\documentclass[prl,reprint,twocolumn,showpacs,superscriptaddress]{revtex4-1}
\usepackage{graphicx}
\usepackage{dcolumn}
\usepackage{bm}
\usepackage{graphicx}
\usepackage{epsfig}
\usepackage{epsf}
\usepackage{amssymb}
\usepackage{amsmath,empheq,cases}
\usepackage{amsthm}
\usepackage{mathtools}
\usepackage{multirow}
\usepackage[colorlinks=true,linkcolor=blue,citecolor=blue,urlcolor=blue,pdfauthor={ },pdftitle={ },pdfsubject={ },pdfkeywords={ }]{hyperref}
\usepackage{pgfplots}
\usepackage{lipsum}

\usepackage{times}

\usepackage{color}
\definecolor{dred}{rgb}{.8,0.2,.2}
\definecolor{ddred}{rgb}{.8,0.5,.5}
\definecolor{dblue}{rgb}{.2,0.2,.8}
\definecolor{dgreen}{rgb}{.2,0.5,.2}

\usepackage{tikz}
\definecolor{c2}{RGB}{153,51,51}

\newcommand{\ket}[1]{|{#1}\rangle}

\newcommand{\bra}[1]{\langle{#1}|}

\usepackage[colorinlistoftodos,bordercolor=red!50,backgroundcolor=red!10,linecolor=red!50,textsize=tiny]{todonotes}
\usepackage{changes}

\newcommand\redsout{\bgroup\markoverwith{\textcolor{red}{\rule[0.5ex]{2pt}{1pt}}}\ULon}

\begin{document}
	\title{Experimental Realization of a Quantum Refrigerator Driven by Indefinite Causal Orders}
	
	\author{Xinfang Nie}
	\thanks{These authors contributed equally to this work.}
	\affiliation{Shenzhen Institute for Quantum Science and Engineering and Department of Physics, Southern University of Science and Technology, Shenzhen 518055, China}
	\affiliation{Guangdong Provincial Key Laboratory of Quantum Science and Engineering, Southern University of Science and Technology, Shenzhen 518055, China}

	\author{Xuanran Zhu}
	\thanks{These authors contributed equally to this work.}
	\affiliation{Shenzhen Institute for Quantum Science and Engineering and Department of Physics, Southern University of Science and Technology, Shenzhen 518055, China}
	\author{Keyi Huang}
	\affiliation{Shenzhen Institute for Quantum Science and Engineering and Department of Physics, Southern University of Science and Technology, Shenzhen 518055, China}

\author{Kai Tang}
	\affiliation{Shenzhen Institute for Quantum Science and Engineering and Department of Physics, Southern University of Science and Technology, Shenzhen 518055, China}

	\author{Xinyue Long}
	\affiliation{Shenzhen Institute for Quantum Science and Engineering and Department of Physics, Southern University of Science and Technology, Shenzhen 518055, China}
	
	\author{Zidong Lin}
	\affiliation{Shenzhen Institute for Quantum Science and Engineering and Department of Physics, Southern University of Science and Technology, Shenzhen 518055, China}
	
	\author{Yu Tian}
	\affiliation{Shenzhen Institute for Quantum Science and Engineering and Department of Physics, Southern University of Science and Technology, Shenzhen 518055, China}
	
	\author{Chudan Qiu}
	\affiliation{Shenzhen Institute for Quantum Science and Engineering and Department of Physics, Southern University of Science and Technology, Shenzhen 518055, China}
\author{Cheng Xi}
	\affiliation{Shenzhen Institute for Quantum Science and Engineering and Department of Physics, Southern University of Science and Technology, Shenzhen 518055, China}	
	\author{Xiaodong Yang}
	\affiliation{Shenzhen Institute for Quantum Science and Engineering and Department of Physics, Southern University of Science and Technology, Shenzhen 518055, China}

	\author{Jun Li}
	\affiliation{Shenzhen Institute for Quantum Science and Engineering and Department of Physics, Southern University of Science and Technology, Shenzhen 518055, China}
	\affiliation{Guangdong Provincial Key Laboratory of Quantum Science and Engineering, Southern University of Science and Technology, Shenzhen 518055, China}
	
\author{Ying Dong}
   \email{yingdong@zhejianglab.edu.cn}
	\affiliation{Research Center for Quantum Sensing, Zhejiang Lab, Hangzhou, Zhejiang, 311121, China}	
	\author{Tao Xin}
	\email{xint@sustech.edu.cn}
	\affiliation{Shenzhen Institute for Quantum Science and Engineering and Department of Physics, Southern University of Science and Technology, Shenzhen 518055, China}
	\affiliation{Guangdong Provincial Key Laboratory of Quantum Science and Engineering, Southern University of Science and Technology, Shenzhen 518055, China}
	
	\author{Dawei Lu}
	\email{ludw@sustech.edu.cn}
	\affiliation{Shenzhen Institute for Quantum Science and Engineering and Department of Physics, Southern University of Science and Technology, Shenzhen 518055, China}
	\affiliation{Guangdong Provincial Key Laboratory of Quantum Science and Engineering, Southern University of Science and Technology, Shenzhen 518055, China}
	\date{\today}
	\begin{abstract}
		
		Indefinite causal order (ICO) is playing a key role in recent quantum technologies. Here, we experimentally study quantum thermodynamics driven by ICO on nuclear spins using the nuclear magnetic resonance system.
		We realize the ICO of two thermalizing channels to exhibit how the mechanism works, and show that the working substance can be cooled or heated albeit it undergoes thermal contacts with reservoirs of the same temperature. Moreover,
		we construct a single cycle of the ICO refrigerator based on the Maxwell's demon mechanism, and evaluate its performance by measuring the work consumption and the heat energy extracted from the low-temperature reservoir. Unlike classical refrigerators in which the coefficient of performance (COP) is perversely higher the closer the temperature of the high-temperature and low-temperature reservoirs are to each other, the ICO refrigerator's COP is always bounded to small values due to the non-unit success probability in projecting the ancillary qubit to the preferable subspace. To enhance the COP, we propose and experimentally demonstrate a general framework based on the density matrix exponentiation (DME) approach, as an extension to the ICO refrigeration. The COP is observed to be enhanced by more than three times with the DME approach. Our work demonstrates a new way for non-classical heat exchange, and paves the way towards construction of quantum refrigerators on a quantum system.    
		
	\end{abstract}
	
	\maketitle
	\emph{Introduction.} --
	Quantum thermodynamics, as an interdisciplinary between thermodynamics and quantum mechanics, aims to broaden the standard thermodynamics to the system dominated by quantum mechanics~\cite{QuantumThermodynamics,kieu2006quantum,PhysRevLett.93.140403,PhysRevE.72.056110,cengel2007thermodynamics,kosloff2013quantum,campisi2011colloquium,esposito2009nonequilibrium}.
	Facilitated by the development of quantum information theory and experimental techniques, quantum heat engine (QHE) has been demonstrated a good platform to study quantum thermodynamics~\cite{maslennikov2019quantum,kondepudi2014modern,dowling2003quantum,micadei2019reversing,PhysRevLett.123.240601,gelbwaser2015thermodynamics,PhysRevA.99.062103,georgescu2012quantum,ribeiro2016quantum,jaeger2018second,PhysRevA.102.012217,zawadzki2020work}.
The idea originates from the Szilard model in 1929 \cite{szilard1929entropieverminderung}, where the famous Maxwell's demon utilizes the information to extract work from the reservoir.
In addition, the usage of quantum resources leads to many counterintuitive effects in QHEs, such as employing squeezed thermal reservoirs to beat the Carnot limit~\cite{PhysRevE.98.042123,PhysRevX.7.031044,PhysRevLett.112.030602} or using effective-negative-temperature reservoirs with nonadiabatic cycles to reach the highest efficiency~\cite{PhysRevLett.122.240602}. Recently, a new counterintuitive thermodynamic resource stemming from the indefinite causal order is shown to be capable of building QHEs \cite{PhysRevLett.125.070603}. Contrary to daily events that happen with definite order, indefinite causal order (ICO) in quantum physics means that
		events can happen without a fixed causal order owing to quantum superposition~\cite{PhysRevA.88.022318,oreshkov2012quantum,rubino2017experimental}.
	ICO is proven to yield superiorities in many research areas, including quantum computation \cite{PhysRevLett.113,PhysRevA.88.022318}, communication complexity \cite{PhysRevLett.122.120504}, metrology \cite{PhysRevLett.124.190503,frey2019indefinite}, and quantum information transmission \cite{PhysRevLett.120.120502,PhysRevLett.124.030502}.
	
Felce and Vedral proposed to apply ICO in quantum thermodynamics~\cite{PhysRevLett.125.070603}. By projecting the ancillary qubit onto its relevant subspace, a working system can be cooled or heated with the ICO of two equivalent thermalizing channels of the same temperature. This is counterintuitive, because empirically exchanging the order of two equivalent channels would not impact on the final temperature.
	This exotic property comes from the magic of quantum information, leading to a new design of quantum refrigerators that only consume work by erasing the information preserved in the ancilla.

	In this Letter, using the ensemble of nuclear spins in the nuclear magnetic resonance (NMR) system~\cite{PhysRevLett.125.090502}, we experimentally demonstrate the ICO process and refrigerators. We show that projective measurements of the ancilla in different basis result in distinct thermodynamic performances. The ICO refrigeration is designed using the Maxwell's demon mechanism to enable the measurement of the average work consumption and heat extraction. The coefficient of performance (COP) is computed based on the experimental data, which is always bounded to small values due to the non-unit success probability in projecting the ancillary qubit to the preferable subspace. To overcome this problem, we extend the ICO refrigeration to a general framework based on density matrix exponentiation (DME) and demonstrate it experimentally, where the COP can be remarkably boosted.
	
\begin{figure*}[t]
		\centering
		\includegraphics[width=0.9\linewidth]{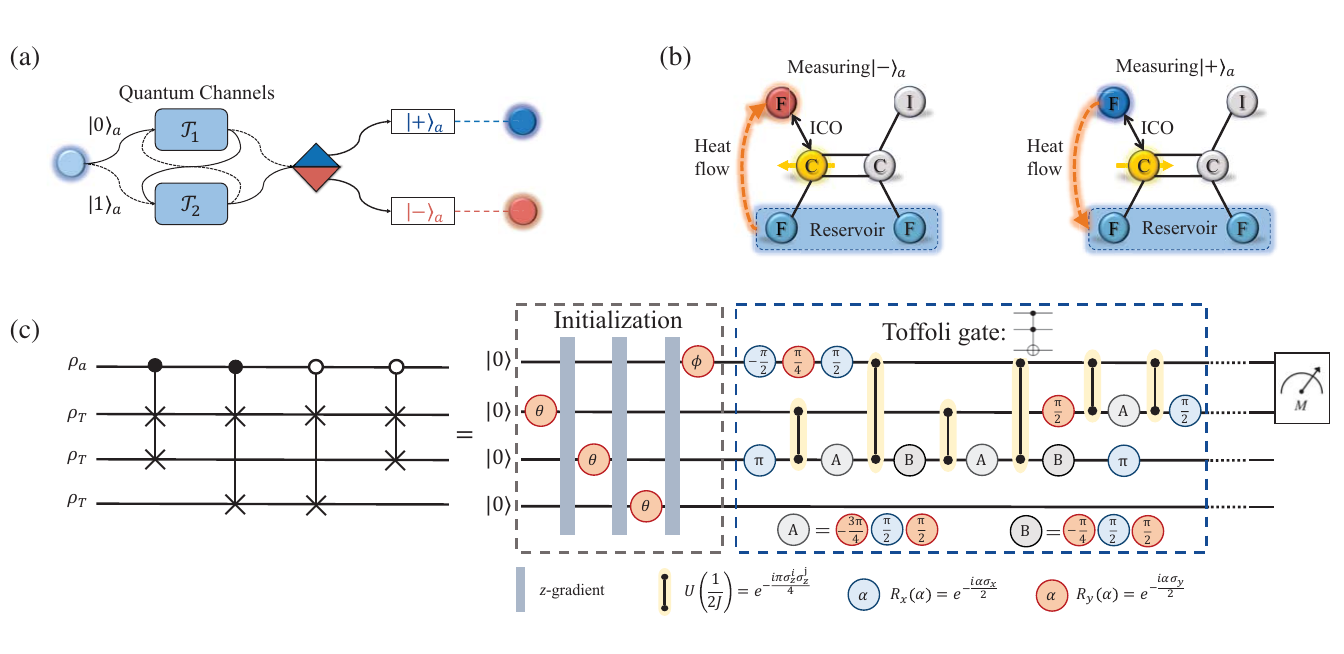}
		\caption{(a) Quantum thermodynamics based on the ICO process. The two channels denoted by $\mathcal{T}_{1}$ and $\mathcal{T}_{2}$ represent two equal thermalizing channels, and the ball represents the working substance. In the beginning, its temperature is the same as that of the two reservoirs. After applying the quantum SWITCH, the ancilla is projected onto its $\vert \pm\rangle_a$ subspace. When $\vert -\rangle_a$ is detected, the heat transfers from the reservoirs to the working substance so that the system is heated (marked by red). Otherwise, the working substance is cooled (marked by blue).
			(b) Implementation of the ICO process on nuclear spins.
			In the sample C$_2$F$_3$I, we use the $^{13}$C as the ancilla, the upper $^{19}$F as the working substance, and two residual $^{19}$F as reservoirs.
			The direction of heat flow depends on the projective measurement result of the ancilla. Explicitly, heat transfers from (to) the reservoir to (from) the working system when the ancilla is measured $\vert-\rangle_a$ ($\vert +\rangle_a$), respectively.
			(c) Quantum circuit and the relevant pulse sequence to implement the quantum SWITCH of thermalizing channels. The first qubit $^{13}$C is the ancilla, the second $^{19}$F is the working substance, and the last two $^{19}$F spins are used to imitate the effect of reservoirs. Each control-SWAP gate can be decomposed into three Toffoli gates. The pulse sequence shows the initialization of the system and the implementation of the first Toffoli gate.
		}
		\label{Fig1}
	\end{figure*}

\emph{Theory.} --
Let us start with a brief review of the ICO thermodynamics. A thermalizing channel of temperature $T$ can transfer an arbitrary input state $\rho$ to a thermal equilibrium state $\rho_{_T}$ at temperature $T$. When two thermalizing channels with the same temperature are applied to the working substance, the final state of the working substance remains the same, regardless of the casual order.
In the quantum realm, channels are allowed to be applied with ICO by introducing an ancilla, whose state controls the casual order of the applied channels.
When the ancilla is initialized in the state $\vert+\rangle_{a}$ and measured in the $\left\{|+\rangle_{a},|-\rangle_{a}\right\}$ basis, the final state of the working system will be
	\begin{equation}
		\rho_\pm = 		\frac{\rho_{_T}\pm\rho_{_T}\rho \rho_{_T}}{2P_\pm},
	\end{equation}
with the probability $P_{\pm}=\text{Tr}[(\rho_{_T}\pm\rho_{_T}\rho \rho_{_T})/2]$.
This means that the effective temperature of the final state can be higher or lower than $T$ depending on the measurement results, as shown in Fig.~\ref{Fig1}(a). Even if we set the initial state $\rho$'s temperature as $T$ -- the same as that of the reservoirs -- the final temperature can still vary. The key ingredient for this counterintuitive phenomenon is that the Kraus operators of thermalizing channels do not commute with each other. This characteristic can be utilized to construct a so-called ICO quantum refrigerator~\cite{PhysRevLett.125.070603} when the working system and the reservoirs are at the same temperature, which is classically impossible. However, the COP of the ICO refrigerator is bounded to a small value due to the low success probability of the measurement.

To enhance the COP, we generalize the ICO refrigerator to a DME~\cite{lloyd2014quantum,PhysRevX.12.011005} driven one.
Note that the controlled-SWAP gates in the ICO quantum circuit in Fig.~\ref{Fig1}(c) is a special form of the controlled-$e^{iS\Delta\theta}$ gate, where $S$ is the standard SWAP operator and $\Delta\theta$ is a tunable parameter \cite{supple}. We modify the ICO circuit by replacing the controlled-SWAP by controlled-$e^{iS\Delta\theta}$, and allow the working system to repeatedly contact the reservoir
for $N_r$ times. This is in fact the way of implementing DME, as proposed in Ref. \cite{lloyd2014quantum} for the purpose of realizing quantum principal component analysis. The original ICO can be considered as a special case in the DME framework with $\Delta\theta=\pi/2$ and $N_r=4$. In the limit of infinite $N_r$ and infinitesimal $\Delta\theta$, the COP can be significantly enhanced compared to that of the ICO refrigeration. We leave the details of the DME framework, including the quantum circuit, analysis of performance, and experimental demonstrations, in the Supplemental Information (SI) \cite{supple}. In the following part of this Letter, we still focus on the realization of the original ICO refrigeration, as it is the most representative example of the DME framework.


	\emph{Realizing the ICO process.} -- The quantum SWITCH channel can be realized by a unitary quantum circuit as shown in the left of Fig.~\ref{Fig1}(c) \cite{PhysRevLett.125.070603}. In this circuit, the first qubit is assigned as the ancilla, the second as the working substance, and the last two in thermal equilibrium state $\rho_{_T}$ as two reservoirs.
	
	In experiment, we use the four nuclei in $^{13}$C-iodotrifluoroethylene (C$_2$F$_3$I)~\cite{PhysRevX.7.031011,PhysRevA.102.012610,PhysRevLett.124.250601} dissolved in acetone-d6 to realize the ICO process, as shown in Fig.~\ref{Fig1}(b). The four nuclei represented by colorful balls are assigned as the four qubits, where the $^{13}$C (yellow), $^{19}$F (red and blue), and two residual $^{19}$F (cyan) are used as the ancilla, working substance, and reservoirs, respectively.
	Experiments are conducted at room temperature on a Bruker AVANCE $600$ MHz NMR spectrometer.
    See the Supplemental Information (SI) \cite{supple} for the information of the sample.
	
	The whole system starts at $\rho_0=\rho_a\otimes\rho_{_T}\otimes\rho_{_T}\otimes\rho_{_T}$ (see SI for the initialization of the system \cite{supple}). Here, we have fixed the initial effective temperature of the working substance as $T$ -- the same as that of the reservoirs. This setting can fully exhibit the intriguing action of the ICO process: even if the working substance and reservoirs have exactly the same starting temperature, the working substance can be warmed up or cooled down. 
	
	
	The quantum SWITCH operation is decomposed by a concatenation of control pulses and free Hamiltonian evolutions in experiment, as shown in Fig. \ref{Fig1}(c). Since each control-SWAP gate can be decomposed into three Toffoli gates ~\cite{nielsen2002quantum}, we just plot the complete pulse sequence of the first Toffoli gate as an example. Then, we utilize the sequence compiler to reduce its complexity and gradient engineering optimization~\cite{khaneja2005optimal} to improve the control accuracy. The final shaped pulse after optimization is $25$ ms with a simulated fidelity above $0.995$. For readout, we perform two-qubit tomography~\cite{PhysRevLett.116.230501,PhysRevLett.118.020401} on the ancilla and work substance \cite{supple}, and reconstruct the experimental output state $\rho_e$. We compute the fidelity between $\rho_e$ and the relevant theoretical prediction for all initial states. The fidelities are always over $0.96$, indicating the well performance of the ICO process.
	
	\begin{figure}[t]
		\centering
		\includegraphics[width=1\linewidth]{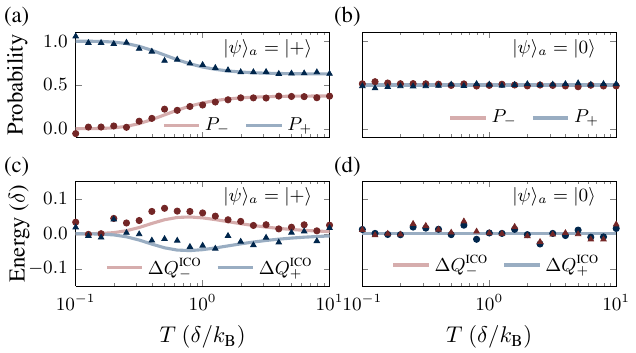}
		\caption{(a-b) Probabilities $P_{\pm}$ of measuring the ancilla in the $\ket{\pm}$ basis. The ancilla is initialized in an (a) equal superposition state $\ket{+}$ and (b) $\vert 0\rangle$, respectively. Triangles and dots are experimental results, and solid lines are theoretical predictions. (c-d)
			Heat transfer between the working substance and reservoirs by measuring the ancilla in the $\ket{\pm}$ basis, when the ancillary qubit is initialized in (c) $\ket{+}$  and (d) $\vert 0\rangle$, respectively. }
		\label{Fig2}
	\end{figure}
	
	\emph{Results of the ICO process.} -- The first quantity from experimental investigation is the probability of projecting the ancilla to $\vert \pm \rangle_a$, which is $P_{\pm}=\operatorname{Tr}({_a\langle\pm\vert \rho_e\vert \pm\rangle_a})$.
	This important quantity reflects the success probability for particular heat flow direction. We fix the ancilla state $\vert \psi_a\rangle$, and vary the temperature $T$ of the working substance and two reservoirs.
	The results are shown in Fig.~\ref{Fig2}(a) and \ref{Fig2}(b), where $\vert \psi_a\rangle$ is $\ket{+}$ and  $\ket{0}$, respectively. Fig.~\ref{Fig2}(a) is the standard ICO process, where $P_{\pm}$ vary against temperature $T$. When temperature goes lower, more remarkable quantum effect emerges as it is more possible to measure $\ket{+}$ on the ancilla, which is related to a cool down of the working substance. For higher temperatures, $P_{+}$ and $P_{-}$ eventually achieve stable values of 0.63 and 0.37, respectively. In Fig.~\ref{Fig2}(b), the input $\vert \psi_a\rangle = \ket{0}$ means a classical situation. Not surprisingly, no differences are observed for measuring the ancilla in the $\ket{\pm}$ basis. For other input states $\ket{\psi_a} = \cos\frac{\phi}{2}\ket{0}+\sin\frac{\phi}{2}\ket{1}$ with unbalanced superpositions, see SI \cite{supple}.
	
	The second important quantity is the amount of heat transfer between the working substance and reservoirs.  In the ICO process, the average heat exchange of the working substance can be defined as
	\begin{equation}\label{Eq3}
		\Delta Q^\text{ICO}_\pm=P_\pm[\operatorname{Tr}(\rho_\pm H)-\operatorname{Tr}(\rho_{_T} H)],
	\end{equation}
	where $\rho_\pm = { _a\langle\pm\vert \rho_e\vert \pm\rangle_a}/{P_{\pm}}$ are the final states of the working substance when the ancilla is projected onto $\vert \pm\rangle$.
	The results  are shown in Fig.~\ref{Fig2}(c) and \ref{Fig2}(d) for the ICO and classical case, respectively.
	As expected, nonzero heat flow only occurs in the ICO process.
	Measuring $\vert+\rangle_a$ indicates heat transfer from the working substance to reservoirs, which can thus be used to construct heat engines. On the contrary, the heat flow reverses which provides resources to construct quantum refrigerators. Due to the conservation of energy, the amount of heat transfer must equal for the heat-up and cool-down processes, \emph{i.e.}, $\Delta Q_+^\text{ICO} + \Delta Q_-^\text{ICO}=0$. Nevertheless, maximal heat flow happens at $T=0.8\delta/k_\text{B}$.

	\emph{Realizing the ICO refrigerator.} -- The ICO process can be used to construct QHEs or refrigerators.
	Here, we demonstrate a quantum cooling cycle driven by ICO in experiment. The COP is studied in terms of the Maxwell's demon mechanism, in which the work consumption and heat extraction can be straightforwardly quantified.
	
\begin{figure}[t]
		\centering
		\includegraphics[width=0.9\linewidth]{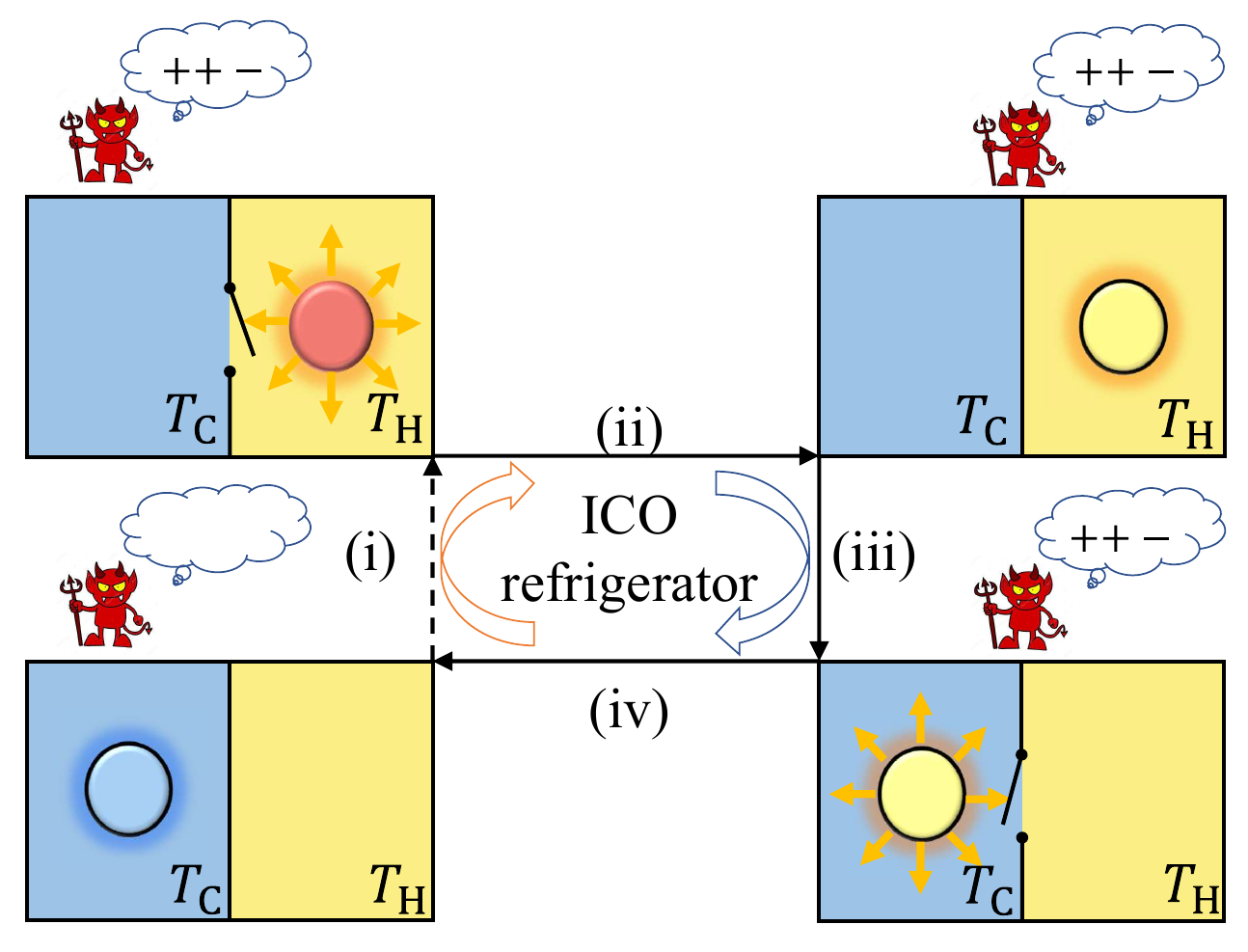}
		\caption{
			Four-stroke ICO refrigerator cycle, which consists of two quantum stages (i) and (iv), and two classical heat exchange stages (ii) and (iii). After the ICO process (i), the Maxwell's demon measures the ancilla. The cycle continues only when the demon reads $\ket{-}_a$. Strokes (ii) and (iii) are thermal contacts with the hot and cold reservoirs, respectively. The last stroke (iv) reinitializes the working substance and erases the demon's memory.
		}.
		\label{Fig3}
	\end{figure}

	An ICO refrigerator consists of four strokes as shown in Fig.~\ref{Fig3}:
	(i) ICO process where projective measurement of the ancilla can be repeated. A Maxwell's demon allows the working substance to continue the cycle if $\vert -\rangle_a$ is measured;
	(ii) classical heat exchange with the high-temperature reservoir and heat rejection; (iii) classical thermal contact with the low-temperature reservoir and heat rejection;
	(iv) initialization of the ancilla and erasure of the Maxwell's demon's memory. To evaluate the COP, we need to measure the heat energy extracted from the cold reservoir to the heat bath $\Delta Q_\text{C}$ (what we want), and the work consumption $W$ (what we pay for) in a cycle.
	
\begin{figure}[t]
		\centering
		\includegraphics[width=0.9\linewidth]{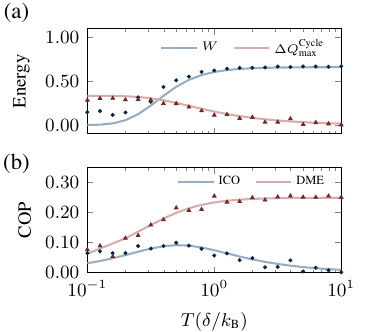}
		\caption{
			(a) Work consumption $W$ during a single ICO refrigeration cycle in theory (blue curve) and experiment (dots), and the heat transfer from the cold reservoir $\Delta Q_\text{C}$ when the ancilla is measured in $\vert -\rangle_a$ in theory (red curve) and experiment (triangles). The hot and cold reservoirs are set at the same temperature $T_\text{H}=T_\text{C}=T$.
			(b) COP of the ICO and DME refrigerations as a function of $T_\text{C}$. The solid lines and dots are theoretical and experimental results of the ICO refrigeration, respectively. The red solid lines and the triangles are the results of the DME framework, which reaches around $\eta_\text{max} = 0.25$ compared to ICO's $\eta_\text{max} = 0.08$.}
		\label{Fig4}
	\end{figure}

	Assuming that all thermalizations are isochoric, the consumption of work only happens in stroke (iv), \emph{i.e.}, when the demon's memory is erased.
	The expected work expenditure to reset the memory in contact with a resetting reservoir of temperature $T_\text{R}$ is $W=k_\text{B}T_\text{R}S$.
	Here, $S=-(P_-\ln P_- +P_+\ln P_+)$ is the Shannon entropy of the demon's memory, where $P_\pm$ is the probability of projecting the ancilla qubit to $\ket{\pm}_a$.
	In experiment, we consider a constant resetting reservoir with $T_\text{R}=\Delta/k_\text{B}$, and the experimental work cost is given in Fig.~\ref{Fig4}(a) by blue dots. Clearly, the work consumption $W$ increases as the cold reservoir's temperature $T_\text{C}$ goes higher.
	The energy flow during strokes (i) - (iii) is in the form of heat exchange. Denoting the heat exchange of the working substance in the $i$-th stroke as  $\Delta Q_i$, the net heat transfer of the cold reservoir in the whole cycle is $\Delta Q_\text{C}=\Delta Q_2=-(\Delta Q_1+\Delta Q_3)$, where the last equality comes from the conservation of the internal energy during the three strokes.  The heat exchange in the second stroke is $\Delta Q_2=\operatorname{Tr}(\rho_- H)-\operatorname{Tr}(\rho_{T_\text{H}} H)$,
	where $\rho_{T_\text{H}}$ is the thermal equilibrium state of the hot reservoir.
	Therefore, to determine the net heat transfer $\Delta Q_\text{C}$ from the cold reservoir, we just need to measure the released heat of the working substance in stroke (ii). This value depends on both the temperature of the hot reservoir $T_\text{H}$ and cold reservoir $T_\text{C}$. It is well known that the COP of a classical Carnot refrigerator approaches infinity when the temperature of the two reservoirs get closer. Here, we study the COP of the ICO refrigeration in an extreme case, \emph{i.e.}, $T_\text{H} = T_\text{C}$.
	At the end of stroke (i), the four-qubit system is prepared into $\ket{+}\bra{+} \otimes \rho_+\otimes\rho_{T_\text{H}} \otimes \rho_{T_\text{C}}$. We set $T_\text{H}=T_\text{C}=T$ as an tunable parameter. Stroke (ii) is a thermal contact between the working substance and the hot reservoir, which can be simulated by a SWAP operation between qubits 2 and 3.  We perform quantum state tomography on the working substance, yielding its internal energy $\operatorname{Tr}(\rho^e_{T_\text{H}}H)$ in experiment. The heat transfer $\Delta Q_\text{C} = \Delta Q_2$ can thus be calculated, combining with the measured $\rho_-$ in the ICO process. The result is displayed with red triangles in Fig.~\ref{Fig4}(a), illustrating that $\Delta Q_\text{C}$ decreases with the increase of $T$.
	
	After measuring the work consumption $W$ and heat transfer from the cold reservoir $\Delta Q_\text{C}$, we evaluate the  COP at different reservoir temperatures.
	Unlike the classical case where the COP is simply $\Delta Q_\text{C}/W$, the ICO refrigerator is conditioned on whether the ancilla collapses to the $\vert -\rangle_a$ subspace. While the work consumption always holds regardless of the measurement results, the success probability $P_-$ must be taking into account when calculating the COP. Hence, the COP of the ICO refrigerator should be defined as
	\begin{equation}\label{Eq5}
		\eta=\frac{\text{what we want}}{\text{what we pay for}}=\frac{\Delta Q_\text{C}}{W/P_-}.
	\end{equation}
	From the results shown in Fig.~\ref{Fig4}(b), we first see that $\eta$ is not infinitely high at $T_\text{H}=T_\text{C}=T$, in contrary to the ideal Carnot refrigerator. Moreover, it varies with the temperature of the reservoir, and reaches an optimal value  $\eta_\text{max}=0.08$ at $T=0.6\delta/k_\text{B}$. This result is genuinely distinct from what we have learned in the classical refrigeration, revealing unique properties of the ICO-based cooling.
	
	\emph{Discussion.} --
As demonstrated above, although the ICO enables heat exchange between the working substance and reservoirs of the same temperature, its COP is even lower than the classical counterpart.  Regarding this issue, we generalize the ICO refrigeration to the DME framework and demonstrate it experimentally, where the COP is significantly enhanced. Notice that the key ingredients in Fig. 1(c) are controlled-SWAP gates, which is a special form of controlled-$e^{iS\Delta\theta}$. Here, $\Delta \theta$ is a variable and $S$ is the standard SWAP operator. We replace the controlled-SWAP gates with controlled-$e^{iS\Delta\theta}$ and allow the working system to repeatedly contact the reservoir for $N_r$ times. When $N_r \rightarrow \infty$, repeated applications of controlled-$e^{iS\Delta\theta}$ is a way of realizing DME, which was proposed in the context of quantum machine learning \cite{lloyd2014quantum} and now broadly applied in quantum computation and metrology. We demonstrate that the DME approach can remarkably improve the COP, as shown in Fig.~\ref{Fig4}(b), where the ICO refrigeration is a special case of the DME framework. We leave technical and experimental details in SI \cite{supple}.

Back to the experiment, we demonstrate that a quantum SWITCH of thermalizing channels supplies a new class of thermodynamic resources. Unlike a deterministic machine whose behavior can be characterized by a single thermodynamic cycle,  the quantum machine driven by the ICO process is however not determinate. Moreover, two counterintuitive phenomena of the ICO process are observed: the working substance can be warmed up or cooled down even if it has the same temperature as that of reservoirs, and the COP is bounded to small values at the extreme case. Compared to previous QHEs with non-equilibrium reservoirs such as squeezed \cite{PhysRevLett.112.030602,PhysRevX.7.031044} or effective-negative-temperature reservoirs \cite{PhysRevLett.122.240602},  the ICO refrigerator does not show superiority in COP. However, a potential advantage of ICO is that the work can be extracted from simpler-to-build equilibrium reservoirs with no temperature difference. The DME framework may shed light on future design of quantum refrigerators with higher COPs, and validates its broad applications in quantum metrology and computation.
	
	\begin{acknowledgments}
		This work is supported by the National Key Research
and Development Program of China (2019YFA0308100),
National Natural Science Foundation of China (12075110,
11975117, 11905099, 11875159, 12104213 and U1801
661), Guangdong Basic and Applied Basic Research
Foundation (2019A1515011383 and 2020A1515110987),
Guangdong International Collaboration Program (2020A
0505100001), Science, Technology and Innovation
Commission of Shenzhen Municipality (ZDSYS20170
303165926217, KQTD20190929173815000, JCYJ20200
109140803865, JCYJ20170412152620376 and JCYJ
20180302174036418), Pengcheng Scholars, Guangdong
Innovative and Entrepreneurial Research Team Program
(2019ZT08C044), Shenzhen Science and Technology
Program (RCYX20200714114522109), Guangdong
Basic and Applied Basic Research Foundation
(2021B1515020070), and Guangdong Provincial Key
Laboratory (2019B121203002). Y. D. is supported by the
Center initiated Research Project of Zhejiang Lab (Grant
No. 2021MB0AL01).
		
		Notes added -- Recently, we become aware of a related
			work by Cao et al., which experimentally studies quantum
			thermodynamics driven by ICO on an optical setup \cite{cao2021experimental}.
	\end{acknowledgments}

%

\end{document}